\documentclass[fleqn,usenatbib]{mnras}

\usepackage{newtxtext,newtxmath}

\usepackage[T1]{fontenc}
\usepackage{ae,aecompl}

\usepackage{graphicx}	
\usepackage{dblfloatfix}
\usepackage{amsmath}	
\usepackage{threeparttable}
\usepackage{subcaption}
\usepackage{caption}
\usepackage{threeparttable}
\usepackage{color}
\usepackage[dvipsnames]{xcolor}
\usepackage[normalem]{ulem}
\usepackage{booktabs}

\usepackage{multirow}
\def \link_col{blue}
\newcommand{\gray}{$\gamma$-ray~}
\newcommand{\grays}{$\gamma$-rays~}

\title[Nuclear line emission from Cas A]{New estimation of the nuclear de-excitation line emission from the supernova remnant Cassiopeia A}
\author[Liu et.al]{Bing Liu$^{1,2,3}$, 
Rui-zhi Yang$^{1,2,3}$\thanks{E-mail: yangrz@ustc.edu.cn}, 
Xin-yu He$^{3,4}$,
Felix Aharonian$^{5,6,7}$
\\
\\
$^{1}$Deep Space Exploration Laboratory/School of Physical Sciences, University of Science and Technology of China, Hefei 230026, China\\
$^{2}$CAS Key Laboratory for Research in Galaxies and Cosmology, Department of Astronomy, School of Physical Sciences, \\
University of Science and Technology of China, Hefei, Anhui 230026, China \\
$^{3}$School of Astronomy and Space Science, University of Science and Technology of China, Hefei, Anhui 230026, China \\
$^{4}$Key Laboratory of Dark Matter and Space Astronomy, Purple Mountain Observatory, 
Chinese Academy of Sciences, Nanjing 210023, China\\
$^{5}$Dublin Institute for Advanced Studies, School of Cosmic Physics, 31 Fitzwilliam Place, Dublin 2, Ireland\\
$^{6}$Max-Planck-Institut f\"ur Kernphysik, Saupfercheckweg 1, 69117 Heidelberg, Germany\\
$^{7}$Yerevan State University,  1 Alek Manukyan St, Yerevan 0025, Armenia\\
}

\date{Accepted XXX. Received YYY; in original form ZZZ}

\pubyear{2021}
\begin{document}
\label{firstpage}
\pagerange{\pageref{firstpage}--\pageref{lastpage}}
\maketitle

\begin{abstract}
MeV nuclear de-excitation lines serve as a unique tool to study low-energy cosmic rays (CRs), containing both spectral and elemental information of the interacting material. In this paper, we estimated the possible nuclear de-excitation lines from the young supernova remnant Cassiopeia A. Given different CR spectral shapes and interacting materials, we found the predicted fluxes of strong narrow line emissions from the remnant are highly model-dependent, ranging from about $1\times10^{-10}\,{\rm \,cm^{-2}\,s^{-1}}$ to $1\times10^{-6}\, {\rm \,cm^{-2}\,s^{-1}}$ for the 4.44 MeV narrow line and from about $4\times10^{-11}\,{\rm \,cm^{-2}\,s^{-1}}$ to  $2\times10^{-7}{\rm \,cm^{-2}\,s^{-1}}$ for the 6.13 MeV narrow line, respectively.  Based on the new estimation, we also discussed the detection probability of these line emissions against the MeV diffuse Galactic background under different assumptions of instrument response functions.   
\end{abstract}

\begin{keywords}
cosmic rays -- gamma-rays: ISM -- ISM: individual objects: Cassiopeia A – ISM: supernova remnants
\end{keywords}

\section{Introduction}
\label{sec:intro}

Cosmic rays (CRs) with kinematic energy below $1~\rm GeV/nucleon$, often referred to as low-energy CRs (LECRs), are most efficient at ionizing and heating gases and play an important role in star-forming and astrochemistry processes \citep{papadopoulos,Gabici2022review}. In the direct measurement of CR spectra in the solar system, the flux of LECRs is highly suppressed by the solar modulation effects. Recently, the Voyager satellite has measured the LECR spectra beyond the heliopause \citep{Cummings2016}. However, it is not straightforward that the LECR spectra measured by Voyager can be a good representative of the LECRs in the Galaxy. Due to the fast cooling and slow propagation of LECRs, their flux indirectly estimated via the ionization rate of gases shows a rather inhomogeneous distribution in the Galactic plane \citep[e.g.,][]{Indriolo2012}.

Supernova remnants (SNRs) are thought to be the most prominent CR accelerators in our Galaxy. The $\gamma$-ray observations in the energy range of $0.1-10~\rm GeV$ from AGILE and Fermi-LAT have shown strong evidence that the SNRs do accelerate CR protons to a high energy range \citep{Giuliani2011,Ackermann2013}. Observations of the large ionization rates from molecular material near SNRs, such as  IC 443, W28, and W49B, suggest that SNRs may also accelerate a large population of LECRs \citep{Indriolo2010,Vaupre2014,Zhou2022}. However, due to the kinetic energy threshold of the pion-decay process ($\sim 280~\rm MeV$), we know little about the injection spectrum of LECRs from SNRs. In addition to the ionization effects, the inelastic collisions between LECRs and interstellar gases could excite the heavy nuclei which can emit MeV \gray lines via de-excitation, such as the 4.44 MeV line from $^{12}$C and the 6.13 MeV line from $^{16}$O \citep[e.g.,][]{Ramaty1979,Murphy2009}. Thus, from observation of these line emissions, we may derive unique information about the injection of LECR nuclei from the accelerators such as SNRs, with the advantage of excluding the influence of CR electrons \citep[e.g.,][]{Benhabiles2013,nlines}.

Cassiopeia A (Cas~A, G111.7-02.1) is the remnant of a massive star explosion $\sim$340 years ago \citep{Fesen2006,krause08}. As one of the youngest SNRs in our Galaxy, Cas~A has been thoroughly investigated from multiwavelength observations despite many open questions that remain debatable.  Located about 3.4 kpc away from the solar system \citep{reed95}, it is one of the brightest sources in the radio band and shows a significant shell structure with an angular radius of 2.5' (or physical size of 2.5 pc) \citep{kassim95}. The synchrotron radiation extends from infrared \citep{tuffs97} to X-rays of about $100~\rm keV$ \citep{casa_nustar}. 
Although the origin of the X-ray radiation is still under debate, \citet{Laming2001a, Laming2001b} argued that non-thermal bremsstrahlung can also explain the observed X-ray flux.  
Early Fermi-LAT observations reveal a hint of the hadronic origin of the \gray emissions \citep{Abdo2010casa,yuan13}. TeV signal from Cas~A was also detected \citep{casa_hegra}, and a significant cutoff at several TeV was revealed by MAGIC and VERITAS observations \citep{Ahnen2017,Abeysekara2020}. Despite the continuous debate about whether Cas~A is a PeVatron or not, a pure hadronic or hybrid origin is preferred to the pure leptonic scenario when explaining the GeV--TeV \gray emission from Cas~A \citep[e.g.,][]{zirak14,Ahnen2017,Zhang2019casa,Abeysekara2020}.  
Thus, one would expect possible MeV de-excitation line emissions arising from Cas~A accelerated LECR nuclei interacting with the surrounding medium.  

In this study, we investigate the potential MeV nuclear line emission from Cas~A under different assumptions of the injected CR spectra which are constrained by recent observations in the GeV-TeV range. Various scenarios of the interacting medium are also considered in the calculations applying the latest estimation of the chemical abundances of the ejecta and ambient gas.  Moreover, the detection capabilities of the line emissions against the continuum background are also discussed regarding the angular resolutions and energy resolutions of the next-generation MeV telescopes.

\section{CR spectra and the medium composition around Cas~A}

Before the calculation of the possible de-excitation $\gamma$-ray line emission from Cas~A, we need to have a general idea of the spectral shape of the accelerated particles and the composition of the interacting medium. Both factors have huge impacts on the estimation results. Given the angular resolution of the next generation MeV telescopes (typically $ \gtrsim 2^\circ$ at MeV band) and the distance of Cas~A, the line emission from Cas~A is very likely to be observed as a point-like source \citep{nlines}. Thus, the spatial distributions of the accelerated LECRs, the interacting medium, as well as the resulting line emission will not be considered in this work.

\subsection{Spectral distribution of the Cas A accelerated particles}
\label{subsec:spectra}
SNRs are widely accepted as one kind of the main CR sources in our Galaxy, and they are expected to accelerate relativistic particles with spectra close to simple power laws in momentum $p$ via the diffusive shock acceleration \citep[e.g.,][]{Bell1978A,Blandford1978}. Given the "test-particle" limit, the CR production rate $q \propto p^{-\chi}$ and the momentum index $\chi\geqslant 2$ in the case of strong shocks.  However, when the nonlinear effects are considered, i.e., the feedback of CR energy and pressure on the shock, the accelerated particles may have spectra that show some concavity in momentum space, and the corresponding low-energy flux will be higher than that of the test-particle predictions \citep[e.g.,][]{Amato2005,Caprioli2011}. 

The energy loss times of the injected protons with kinetic energy $E$ of 1\,MeV and 10\,MeV are about $2\times10^{3}$ yrs and $4\times10^{4}$ yrs assuming an average medium density of $10\,{\rm cm}^{-3}$. Given the relatively young age of Cas~A, the deformation of the CR spectra from the freshly injected spectra of Cas A could be ignored.  Thus, for simplicity, we chose a piece-wise power law in proton momentum $p(E)$ with an exponential cutoff at $E_{\rm cut}$ to describe the spectral distribution of CR protons. The injection flux $F(E)$ is given by 
\begin{equation}
F(E)\!\! =\!\! \left\{ \begin{array}{ll}
N_0 \left[\frac{p(E)}{p(E_{\rm b})}\right]^{-\alpha_1} \exp\left[\frac{-p(E)}{p(E_{\rm cut})}\right],
\; & {\rm if}\, E \geq E_{\rm b}  \\
\\
N_0\left[\frac{p(E)}{p(E_{\rm b})}\right]^{-\alpha_2} \exp\left[\frac{-p(E)}{p(E_{\rm cut})}\right],
\; & {\rm if}\, E < E_{\rm b}  \\
\end{array}\right.\!\!\!.
\label{eq1}
\end{equation}

Here, the cutoff energy $E_{\rm cut}=10$\,TeV and the broken energy $E_b$ is set to be 0.2 GeV, below which the protons do not participate in the pion production process and lose their energy mainly via the ionization process.

\subsection{Elemental abundances of the interacting medium}
\label{sec:ele}

According to multiwavelength observations, the ejecta of Cas~A is dominated by oxygen   ($\sim 2.55 {\rm M}_{\odot}$), and the rest is comprised mainly of Ne, Si, S, Ar, and Fe, with a total mass of about 3--3.5 ${\rm M}_{\odot}$,
meanwhile, the composition of the circumstellar medium (CSM) shows enhancement of N and He relative to the solar abundances and the mass of the shocked CSM is about 10 ${\rm M}_{\odot}$ \citep[e.g.,][]{Chevalier1978,Chevalier1979,Docenko2010,Hwang2012,Laming2020}. 
The ejecta abundances vary with position in the remnant. For example, the study of the X-ray-emitting ejecta in Cas~A using the {\it Chandra} 1 Ms observation see sub-solar ratios \citep{Hwang2012}, while the study of “bulk” unshocked ejecta using the IR Spitzer data get super-solar ratios \citep{Laming2020}. 
To carry on our estimation, regardless of the very large uncertain remains for the exact compositions of the ejecta, we refer to the "average" number density ratios (relative to O) summarized in Table 9 from \citet{Docenko2010} for Ne, Mg, Si, S, Ar, and Fe, and adopt the solar value (relative to O) for C, N, and Ca.  In addition, we change the density ratios of H and He, from 0.01 to 0.05 and 0.1 to 0.5 respectively, to check the impact of their uncertain densities in the ejecta on the flux of the line emission.

For the CSM, same as \citet{Hwang2012}, we set the abundance of He 3 times the solar value, that of N 15 times the solar value, and solar values for the rest. Meanwhile, we apply the Voyager measurement of local LECRs \citep[see][Table~3]{Cummings2016} as a simplified assumption of the abundance of Cas~A accelerated CRs. Above elemental compositions assumed for the calculation are summarized in Table~\ref{tab:ab}. 

We note that a thorough investigation into the composition of Cas~A accelerated particles requires comprehensive knowledge of the acceleration and escape of the particles, as well as the diffusion and mixing of them into the surrounding medium. Such kind of research is beyond the scope of this work, which only serves as an estimate that awaits valuable information from next-generation  MeV \gray detectors.

\begin{table}
\caption{The elemental compositions assumed for calculation}
\begin{tabular}{ccccc}
\hline
 & CR $^{\rm a}$ & Solar $^{\rm b}$ & CSM $^{\rm c}$ &Ejecta $^{\rm d}$ \\
 &$n_{\rm el}/n_{\rm H}$ & $n_{\rm el}/n_{\rm H}$& $n_{\rm el}/n_{\odot}$&$n_{\rm el}/n_{\rm O}$ \\
 \hline
H&  1 & 1& 1& 0.01-0.05$^\ast$ \\
He&  $8.140\times10^{-2}$&$8.414\times10^{-2}$ &3 &0.1-0.5$^{\rm e}$\\ 
C &  $1.671\times10^{-3}$&$2.455\times10^{-4}$&1& 0.5$^{\rm e}$\\ 
N &  $2.444\times10^{-4}$&$7.244\times10^{-5}$&15&0.1$^{\rm e}$\\ 
O &  $1.570\times10^{-3}$&$5.370\times10^{-4}$&1&1  \\ 
Ne &  $1.507\times10^{-4}$&$1.122\times10^{-4}$&1&0.02 \\
Mg&  $2.264\times10^{-4}$& $3.467\times10^{-5}$&1 & 0.005 \\
Si&  $1.898\times10^{-4}$& $3.388\times10^{-5}$&1&0.05  \\
S& $2.087\times10^{-5}$& $1.445\times10^{-5}$&1&0.05  \\
Ar&$4.554\times10^{-6}$& $3.162\times10^{-6}$&1&0.005  \\
Ca&$1.195\times10^{-5}$&  $2.042\times10^{-6}$&1&0.004$^{\rm e}$  \\
Fe&$1.152\times10^{-4}$&$2.884\times10^{-5}$&1&0.005\\
\hline
\end{tabular}
\\
\footnotesize{
$^{\rm a}$ Number density ratio relative to H, adopted from the Voyager measurement of local LECR abundance \citep[see][Table~3]{Cummings2016}.\\
$^{\rm b}$ Number density ratio  relative to H, adopted  the recommended present-day solar abundance \citep[see][Table~6]{Lodders2010}.\\
$^{\rm c}$Number density ratio relative to the solar value, adopted for the CSM of Cas A, mainly referred to \citet{Hwang2012}.\\
$^{\rm d}$ Number density ratio relative to O, mainly adopted from \citet{Docenko2010}.\\
$^{\rm e}$ The details for the settings of these elements are described in Sec.\ref{sec:ele}. \\}
\label{tab:ab}
\end{table}

\section{Possible de-excitation line emission from Cas~A }
\label{sec:nline}
For the calculation of the  $\gamma$-ray line emissions, except for the code TALYS which is the newest 1.96 version \citep{talys2008,Koning2014}, we applied the same procedure as described in Section 3.1 of \citet{nlines}, which followed the method developed by \citet{Ramaty1979,Murphy2009,Benhabiles2013}. Two main channels of the interactions are considered here. One is the {\sl direct} process in which the CR protons and $\alpha$-particles as projectiles excite heavier elements of the ambient gas and generate narrow \gray lines, and the other is the {\sl inverse} process in which the hydrogen and helium of the ambient gas excite the heavy nuclei of LECRs and produce \gray line emission that is broadened.

Given the large difference in the elemental compositions between the ejecta and the CSM,  here we consider three cases to study the possible influence of various interacting materials. One case (hereafter referred to as case 1) is that the \gray emission is generated from the shock-accelerated nuclei interacting only with the ejecta. The second one (case 2) is that the emission is produced by the accelerated nuclei colliding with the CSM. Moreover, a hybrid case (case 3) in which half of the accelerated CRs interact with the ejecta while the other half of the CRs interact with the CSM is also calculated. An estimation of the \gray lines from the same population of CRs colliding with the gas medium of solar abundance is also made for further comparison.  

\begin{figure}
\includegraphics[width=0.4\textwidth]{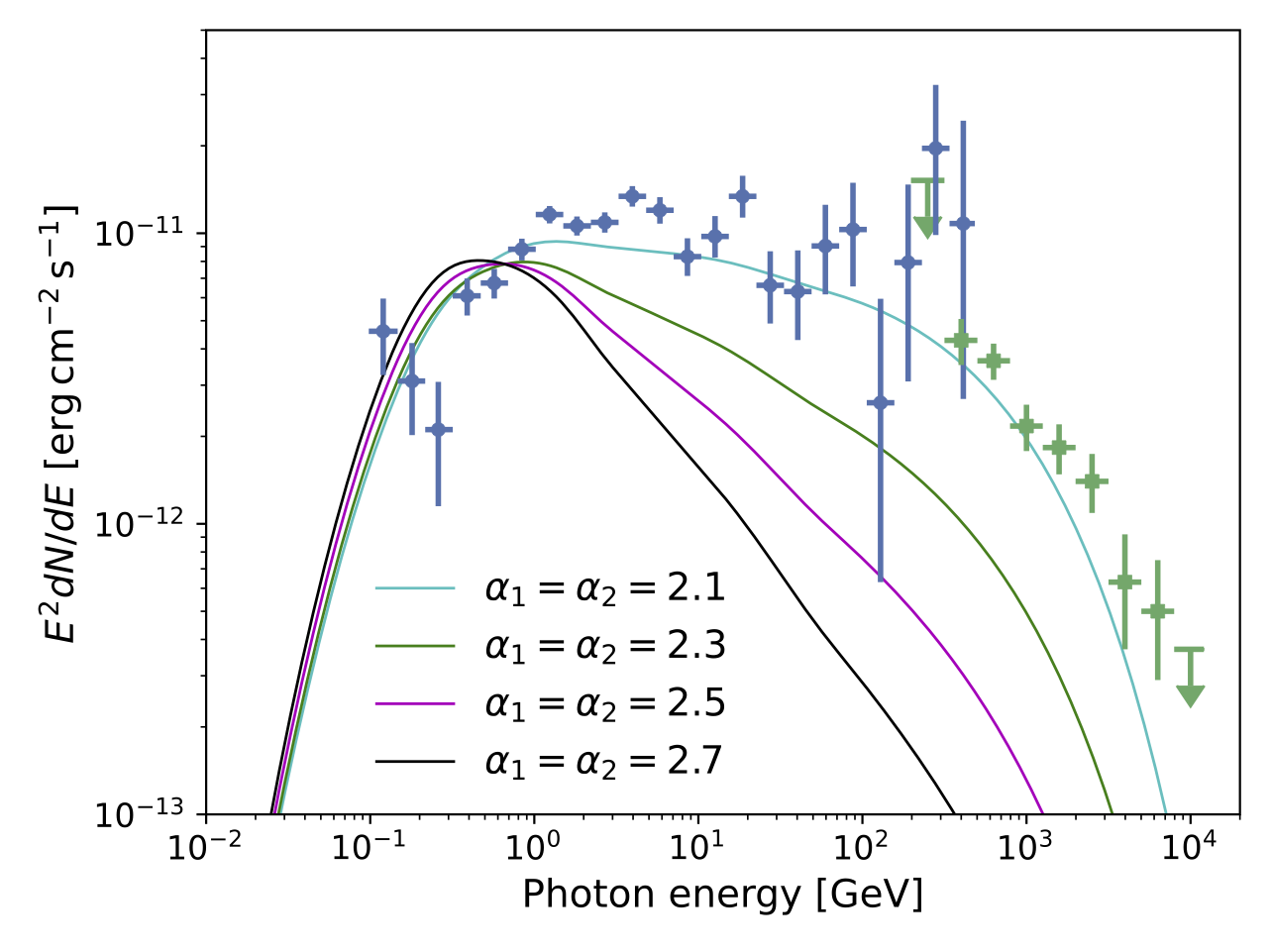}
\caption {Examples of the \gray emission produced via pion-decay process from Cas A with various assumptions of $\alpha_1$ and $\alpha_2$. The data points are adopted from the GeV-TeV \gray observations of \citet{Abeysekara2020}. Details are described in Sec.\ref{subsec:spectra}.} 
\label{fig:cr}
\end{figure}

For each case, we tested possible proton spectra with different assumptions of $\alpha_1$ ranging from 2.0 to 2.7 while setting $\alpha_2=\alpha_1$, 3.0, and 4.0, respectively.  Meanwhile, the GeV-TeV \gray spectral data from recent research of \citet{Abeysekara2020} are used to constrain the overall flux of the accelerated particles, which provides us a maximum for $N_0$ given a certain density and composition of the interacting medium. Here we applied the Eq.(20) in \citet{Kafexhiu2014} to calculate the nuclear enhancement factor $\epsilon$, then derived the effective density $\epsilon n_{\rm p} n_{\rm H}$ for the pion-decay process, in which $n_{\rm p}$ represents the density of the accelerated protons and $n_{\rm H}$ is hydrogen density of the interacting medium.  Examples of the pion-decay emissions derived from the above spectral parameter settings are shown in Fig.\ref{fig:cr}.  

\begin{figure}
\includegraphics[width=0.45\textwidth]{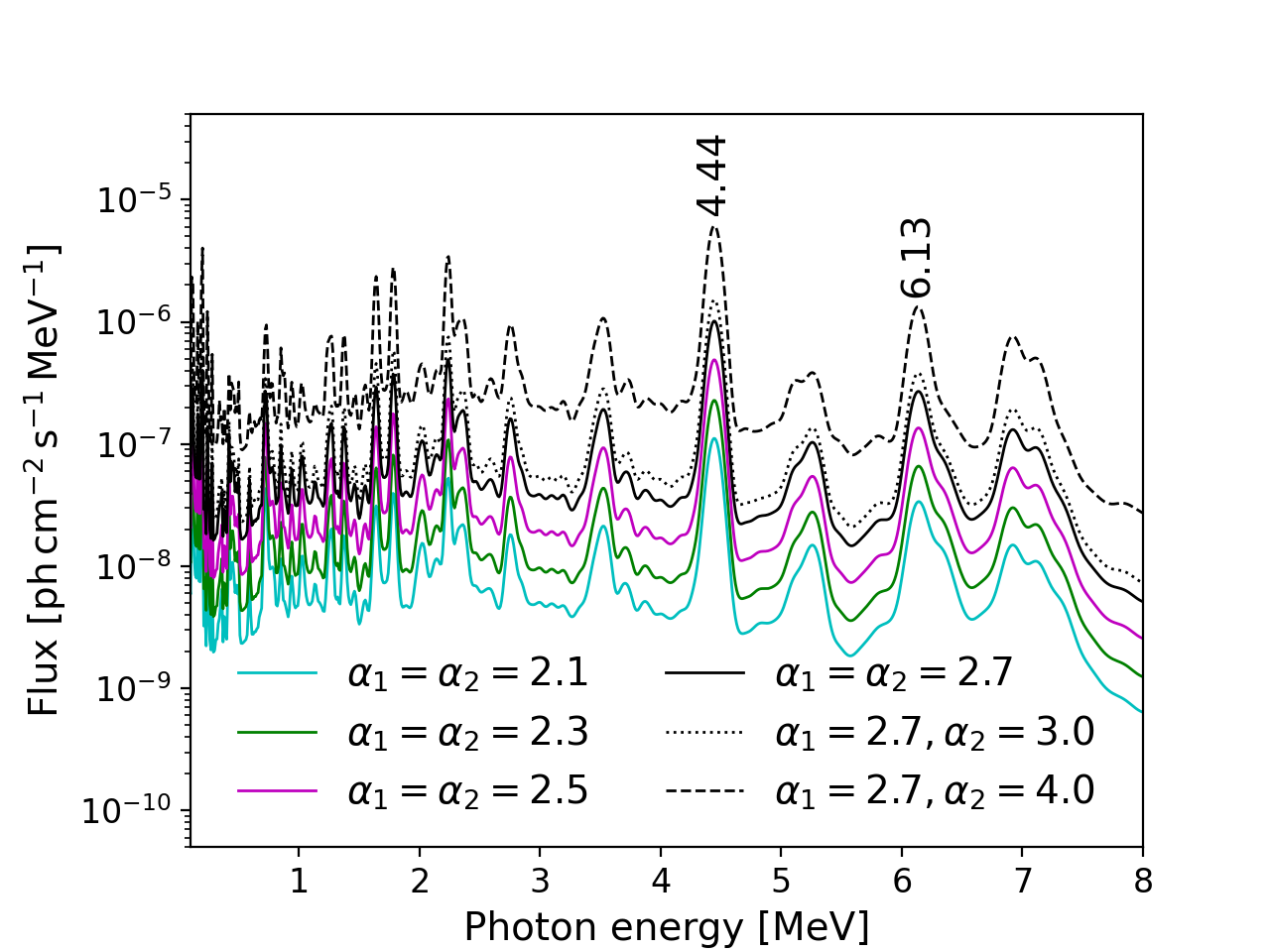}
\caption {Comparison of estimated MeV \gray differential spectra  with different spectral settings of $\alpha_1$ and $\alpha_2$ for case~1 as described in Sec.\ref{sec:nline}. }
\label{fig:case1}
\end{figure}

\begin{figure}
\includegraphics[width=0.45\textwidth]{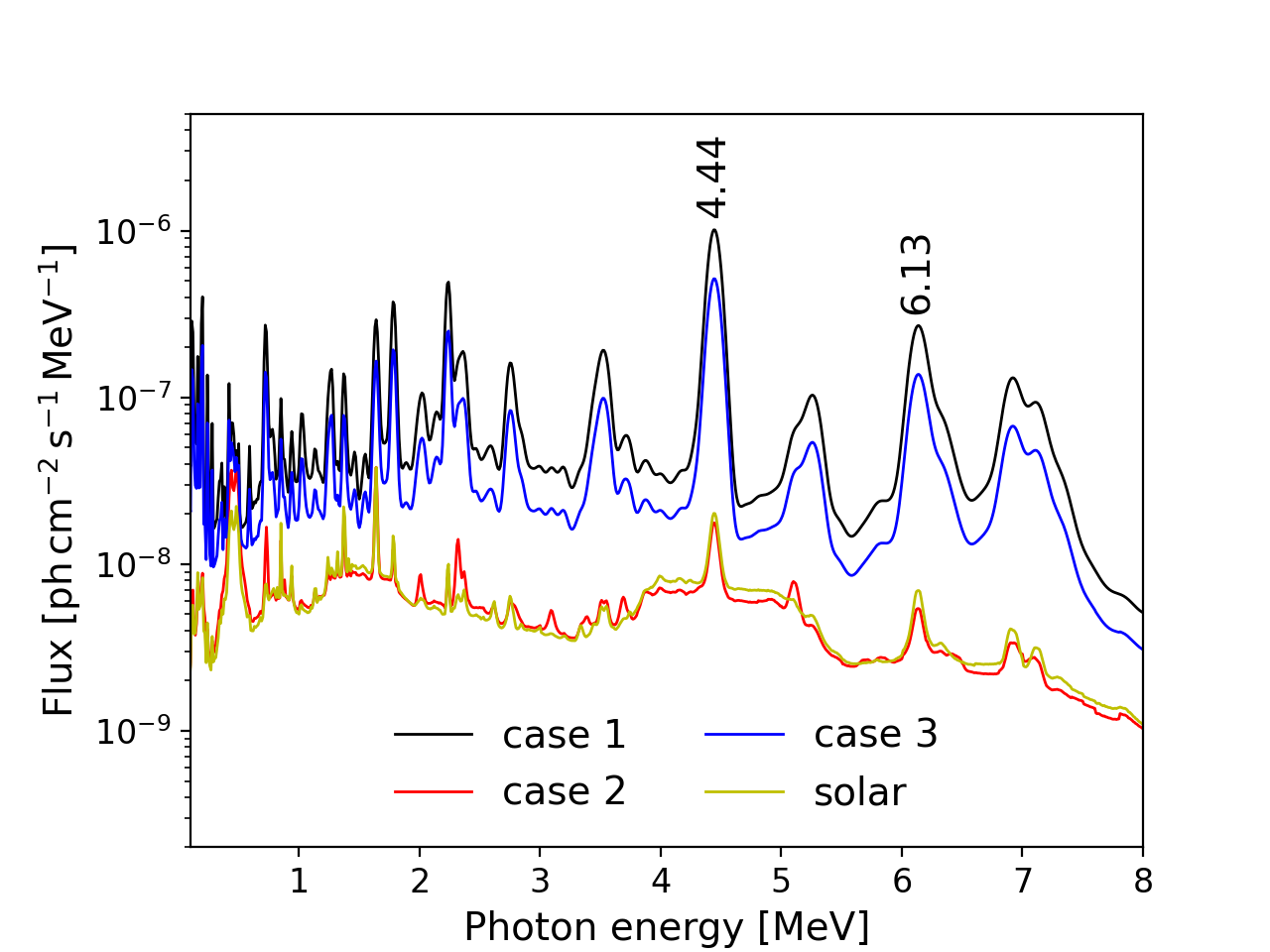}
\caption{Comparison of estimated MeV \gray differential spectra of Cas A with $\alpha_1=\alpha_2=2.7$ for different cases described in Sec.\ref{sec:nline}.}
\label{fig:3cases}
\end{figure}

Considering the uncertainty of the relative density of H and He in the ejecta, we varied their number ratios from 0.01 to 0.05 and from 0.1 to 0.5, respectively, for the calculation of case~1. We found that such changes have very little impact ($\lesssim3\%$) on the overall line flux under the premise of oxygen domination. For a more realistic estimation,  we also considered the Doppler effect caused by the movement of the ejecta \citep[e.g.,][]{Milisavljevic2013}, 
and adopted an $\Delta V$ of $\sim 6000 ~{\rm km\,s}^{-1}$ from the recent study of \citet{Picquenot2021} for all the elements. The resulting MeV nuclear de-excitation line emissions for case~1 are exemplified in Fig.~\ref{fig:case1}. As shown in Fig.~\ref{fig:case1}, the line fluxes increase with the spectral index and the presence of a concavity in momentum space will lead to much stronger line emissions. In addition, the same trend is also found for case~2 and case~3. However, assuming the same CR spectral shape, the line fluxes of case~2, are much lower compared to that of case~1, and the morphological difference of the MeV \gray emission is also very obvious, as illustrated by the solid lines in Fig.~\ref{fig:3cases}. Such contrast reflects the much more abundant heavier nuclei in the oxygen-dominated ejecta compared to those in the CSM.
Taking the possible Doppler broadening into account, for case 1 and case 3, the FWHM widths ($\Delta E$) of the 4.44 MeV line and the 6.13 MeV line are $\sim 0.19$ MeV and $\sim 0.22$ MeV, respectively.  
The integrated narrow-line fluxes of the 4.44 MeV and the 6.13 MeV lines of these three cases are summarized in Table~\ref{tab:lflux}, in which the minimum and maximum are obtained when setting $\alpha_1=\alpha_2=2.0$ and $\alpha_1=2.7$, $\alpha_2=4.0$, respectively. As shown in Table~\ref{tab:lflux}, the integrated line flux ranges are  $\sim (1\times10^{-10}-1\times10^{-6}) {\rm \,cm^{-2}\,s^{-1}}$  for the 4.44 MeV narrow line and $\sim (4\times10^{-11}-2\times10^{-7}){\rm \,cm^{-2}\,s^{-1}}$ for the 6.13 MeV narrow line, respectively. 

Our estimation of the integrated 4.44 MeV line flux assuming $\alpha_1=\alpha_2=2.1$ for case~1 is $\sim 1.4 \times10^{-8} {\rm \,cm^{-2}\,s^{-1}}$, much lower compared to the estimation of \citet{Summa2011casA}. 
Indeed, the target density, composition of the medium, and the LECR flux can vary dramatically with the distance to the forward shock.  Using a uniform density and composition can be quite biased in estimating the MeV line emission. This is also the motivation that we consider three cases in the discussions above.

\begin{table}
\caption{Integrated 4.44 MeV and 6.13 MeV line fluxes of various cases $^{\rm a}$  } 
\begin{tabular}{cccc}
\hline
 & Medium  &4.44 MeV & 6.13 MeV   \\
\hline
case~1& Ejecta &(0.09--7.71)$\times10^{-7}$ &(0.04--2.10)$\times10^{-7}$\\
case~2&  CSM &(0.10--8.84)$\times10^{-9}$&(0.04--2.66)$\times10^{-9}$ \\
case~3& Ejecta+CSM  & (0.05--3.93)$\times10^{-7}$&(0.02--1.06)$\times10^{-7}$ \\
\hline
\end{tabular}
\\
\footnotesize{
$^{\rm a}$ The flux ranges (in units of ${\rm \,cm^{-2}\,s^{-1}}$) are obtained from different assumptions of CR spectral index ($\alpha_1$ and $\alpha_2$). Details are described in Sec.\ref{sec:nline}. \\
}
\label{tab:lflux}
\end{table}

\section{Discussion and Conclusion}
\label{sec:dis}

Regardless of the uncertainties from the experiment data and the simulation data from TALYS, as shown in Sect.\ref{sec:nline}, the de-excitation line fluxes resulting from the interaction between the Cas~A accelerated CR nuclei and the medium are highly model-dependent:
for certain cases, the predicted line fluxes vary by two orders of magnitude due to different settings of the spectral indexes, meanwhile, the narrow line fluxes could also differ by two orders of magnitude due to the variation in the elemental compositions of the interacting medium.

\begin{figure*}
\centering
\includegraphics[width=0.48\linewidth]{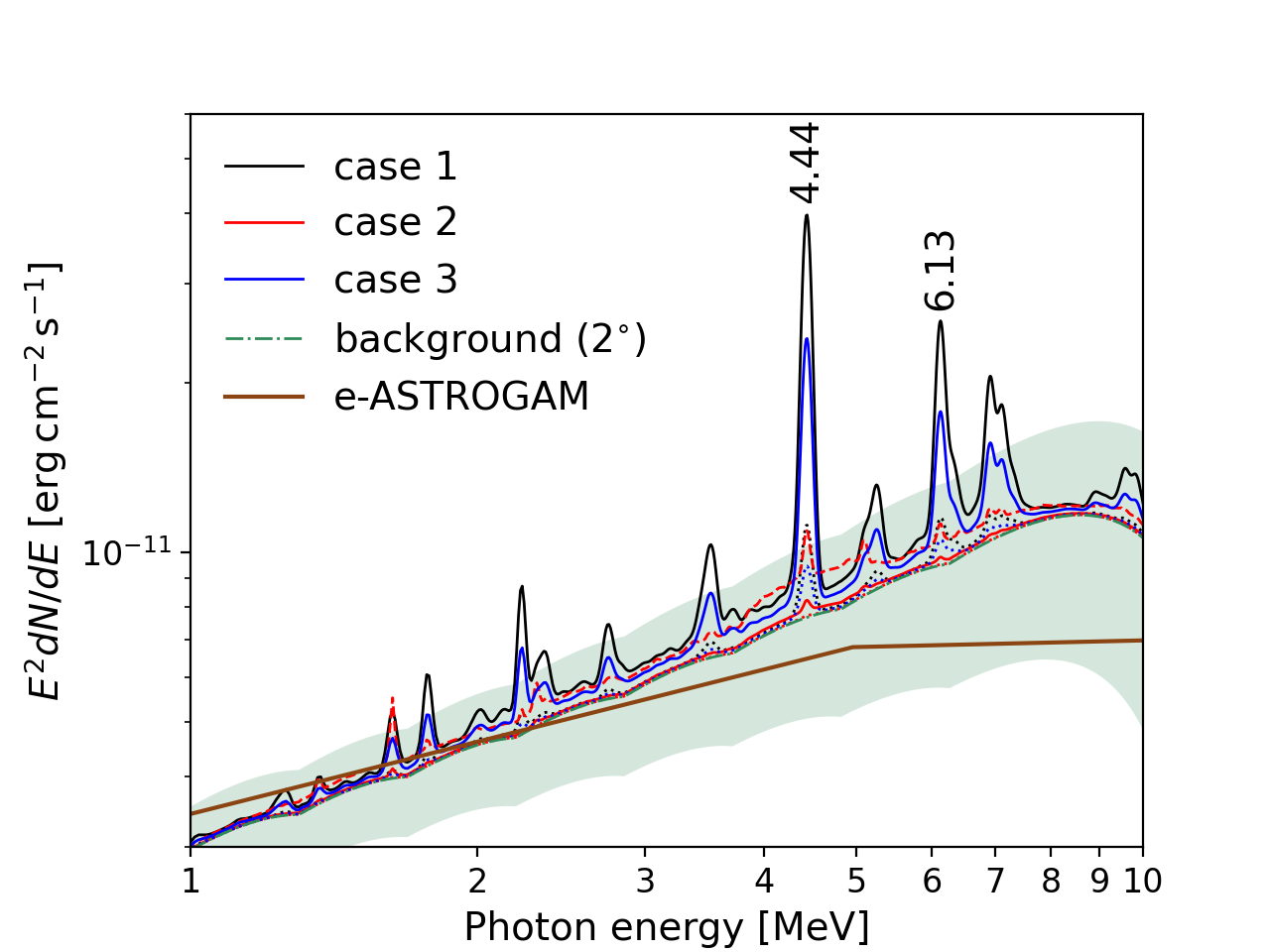}
\includegraphics[width=0.48\linewidth]{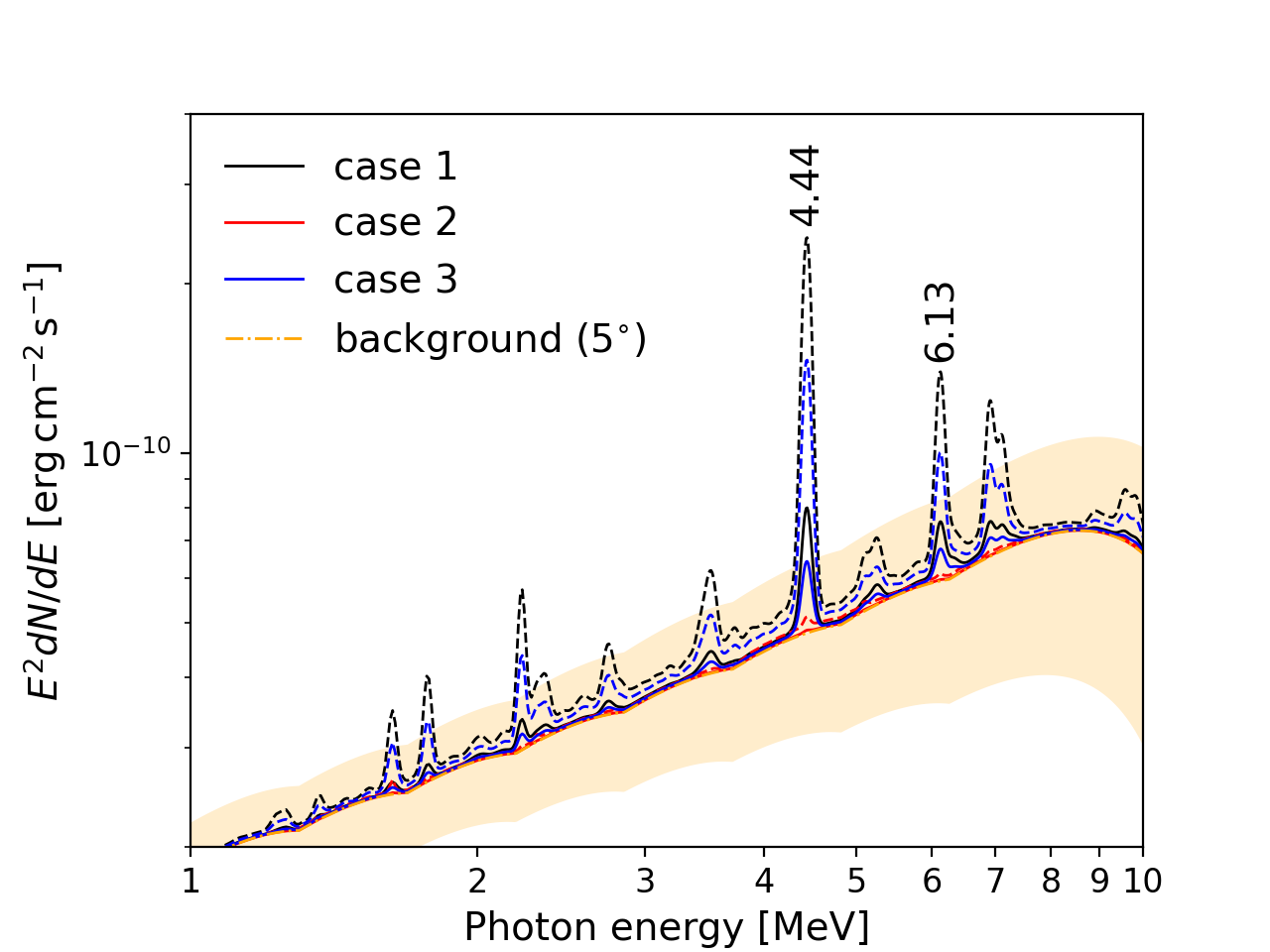}
\caption{The overall MeV \gray emission from Cas A region with extrapolated diffuse background added for various cases and spectral shapes, in which $\alpha_1=\alpha_2=2.7$ is denoted by the solid lines, $\alpha_1=\alpha_2=2.1$ is denoted by the dotted lines, and $\alpha_1=2.7$,$\alpha_2=4.0$ is denoted by the dashed lines. 
 The dash-dotted line and shaded area represent possible diffuse background emission with  $1\sigma$ uncertainty around Cas~A, assuming the angular resolutions of the telescope are 2$^{\circ}$ (green) or 5$^{\circ}$ (orange), respectively. The sensitivity of e-ASTROGAM calculated at 3$\sigma$ for an effective exposure of 1 year and for a source at high Galactic latitude is shown by the brown line \citep{e-astrogam2018}.
See Sect.~\ref{sec:dis} for the details. }
\label{fig:sed1}
\end{figure*}

\begin{figure}
\includegraphics[width=0.48\textwidth]{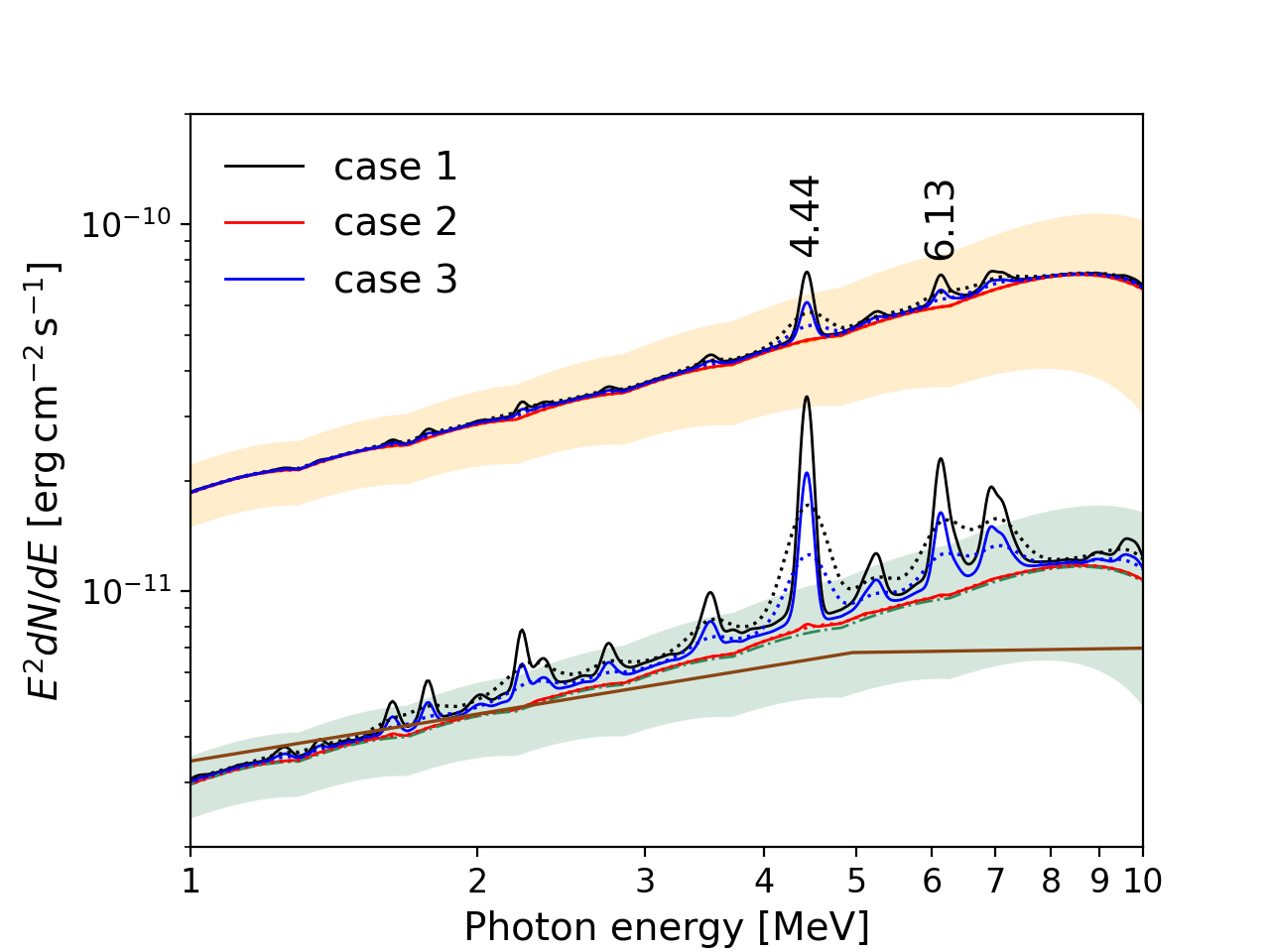 }
\caption{The overall MeV \gray emission from Cas A region with extrapolated diffuse background added  when considering different instrument energy resolutions ($\Delta E/E$) for $\alpha_1=\alpha_2=2.7$.  The assumed energy resolutions ($\Delta E/E$) are $2\%$ for solid lines and $10\%$ for dotted lines, respectively.  The dash-dotted lines, the shaded areas, and the brown line are the same as Fig.~\ref{fig:sed1}.
}
\label{fig:sed2}
\end{figure}

The continuum MeV \grays contributed from the diffuse Galactic emission and Cas A itself should be taken into account when discussing the detectability of line emission. Based on the observation of SPI aboard INTEGRAL, \citet{Siegert2022} re-analyzed the diffuse Galactic emission at 0.5 and 8.0 MeV by fitting energy-dependent spatial template GALPROP \citep{galprop} models within a region of $\Delta l \times \Delta b = 95^{\circ} \times 95^{\circ}$ around the Galactic center. They found the above observed diffuse background is mainly contributed by IC scattering of CR electrons onto the interstellar radiation field, of which the bremsstrahlung component may account for $\sim10\%$.
We estimate the diffuse MeV emission flux in the direction of Cas~A by extrapolating this newest measurement spatially.  The possible diffuse background flux in the direction of Cas~A within 1 $\sigma$ uncertainty is shown as the shaded area in Fig.~\ref{fig:sed1} and Fig.~\ref{fig:sed2}, in which the angular resolutions of the telescope are assumed to be 2$^\circ$ and  5$^\circ$, respectively. As for the continuum MeV emission from Cas~A itself, which is mainly contributed from bremsstrahlung in the hadronic scenarios, the modeled fluxes are well below or at the same level of the extrapolated diffuse background \citep[e.g.,][]{Zhang2019casa,Abeysekara2020}. Thus, the potential influence from the bremsstrahlung on the detection of the line emissions  would be much weaker or similar to that of the diffuse background, and for simplicity, this was not further discussed in the following discussion.

Due to the highly model-dependent calculation of the possible nuclear \gray line emission from Cas~A, future observations from the next-generation MeV telescopes may help us locate the interacting region(s) and constrain the injected CR spectral shape, which can be served as a diagnosis on particle acceleration and escape in Cas A.  
To be specific, if the interacting material is dominated by ejecta, the strong narrow line emission such as the 4.44 MeV line and the 6.13 MeV line are more likely to be detected. 
Moreover, the existence of a concavity in the CR injection spectra (as exemplified by the dash-dot lines in Fig.~\ref{fig:sed1}), would make the detection easier for case~1 (black) and case~3 (blue), even if the flux of the background continuum is very high due to the limited angular resolution. And a better angular resolution can increase the possibility of detection for case~2 (red).  In general,  with the possible diffuse background emission added, the detection of the MeV line emission will be more likely from instruments with better angular resolutions ($~2.0^{\circ}$) for the point-like source SNR Cas~A.  
In addition, we checked the influence on the detectability regarding the energy resolutions ($\Delta  E/E$, FWHM) of the telescopes. The results are shown in Fig.~\ref{fig:sed2}, assuming a $\Delta  E/E$ of $2\%$ (solid lines) or $10\%$ (dotted lines), respectively. We found that the energy resolution is crucial for the detection of such line features. However, as calculated in the section above, we found that the narrow de-excitation lines have intrinsic Doppler broadening of about $2\%$ that is caused by the recoil of the excited nuclei, and additional broadening of about $2\%$ due to the movements of the ejecta-dominated medium.  Thus the energy resolution better than this value can hardly improve the detection sensitivity.  

In conclusion, we estimated the possible MeV line emission from LECRs interaction with the ambient gas in the SNR Cas A under simplified conditions. We found that if the accelerated CRs mainly interact with the ejecta, the line signal from Cas~A will be more prominent due to the higher ratio of heavy nuclei therein. And the potential softening of LECR spectra caused by CR feedback to the shock would further enhance the MeV line emissions. We also found that the diffuse MeV \gray emissions in the Galactic plane may be the main background in the detection of such line features. Both angular resolution and energy resolution play a significant role in detecting these MeV lines from Cas A.  Based on our model-dependent predication, the integrated narrow line fluxes are  $\sim 1\times10^{-10}{\rm \,cm^{-2}\,s^{-1}}$ to $1\times10^{-6} {\rm \,cm^{-2}\,s^{-1}}$ at 4.44 MeV and from $\sim 4\times10^{-11}{\rm \,cm^{-2}\,s^{-1}}$ to $2\times10^{-7}{\rm \,cm^{-2}\,s^{-1}}$ at 6.13 MeV, respectively. Thus, the next generation MeV instruments with a line flux sensitivity of about  $10^{-6}~\rm cm^{-2}s^{-1}$, such as e-ASTROGAM, AMEGO, and COSI  \citep{e-astrogam2018,amego2019,COSI} may have chances of detecting these unique spectral features in Cas A, but detectors with sensitivities exceeding the capability of projects mentioned above would be more promising for the research of such kind of individual CR sources. The recently proposed large-scale Space Projects like MeGaT (Zhang et al. 2023, private communication) and  MeVGRO (Peng et.al 2023, private communication) give optimism that the probes of low-energy particles from Cas A could be realized through the detection of prompt nuclear de-excitation line emission.

\section{Acknowledgements}

Bing Liu acknowledges the support from the NSFC under grant 12103049. Rui-zhi Yang is supported by the NSFC under grants 12041305, and the national youth thousand talents program in China.

\section{Data Availability}

To calculate emissivities of the de-excitation \gray line lines,  we used the code TALYS  (version 1.96,\citet{talys2008}), which could be downloaded from {\url{https://tendl.web.psi.ch/tendl_2019/talys.html}. For a better match with the experiment data, we modified the deformation files of $^{14}$N, $^{20}$Ne, and $^{28}$Si using the results of \citet{Benhabiles2011}. 
We also used the production cross sections of the specific lines listed in the compilation of \citet{Murphy2009}. 



\bibliographystyle{mnras}
\bibliography{cite} 








\bsp
\label{lastpage}

\end{document}